\documentclass[aps,amssymb,prb,showpacs,twocolumn]{revtex4}

\usepackage{epsf}
\usepackage{amsmath}
\usepackage{graphicx}
   
\begin{document}                                                       
                                            
\title{Tunneling in a magnetic field}

\author{B. Ivlev} 

\affiliation
{Department of Physics and Astronomy and NanoCenter\\
University of South Carolina, Columbia, SC 29208\\
and\\
Instituto de F\'{\i}sica, Universidad Aut\'onoma de San Luis Potos\'{\i}\\
San Luis Potos\'{\i}, S. L. P. 78000 Mexico}


\begin{abstract}
Quantum tunneling between two potential wells in a magnetic field can be strongly increased when the potential barrier 
varies in the direction perpendicular to the line connecting the two wells and remains constant along this line. An 
oscillatory structure of the wave function is formed in the direction joining the wells. The resulting motion can be 
coherent like motion in a conventional narrow band periodic structure. A particle penetrates the barrier over a long 
distance which strongly contrasts to WKB-like tunneling. The whole problem is stationary. A not very small tunneling 
transparency can be set between two quantum wires with real physical parameters and separated by a long potential barrier.
The phenomenon is connected to Euclidean resonance. 

\end{abstract} \vskip 1.0cm
   
\pacs{03.65.Xp, 03.65.Sq} 
 
\maketitle

\section{INTRODUCTION}
According to Wentzel, Kramers, and Brillouin (WKB) [\cite{LANDAU}], there is a finite probability $w\sim\exp(-A)$ of 
quantum tunneling through an one-dimensional potential barrier. This probability becomes negligible for semiclassical 
barriers when $A=2{\rm Im}S/\hbar$ and a classical under-barrier action $S$ is big. In two-dimensions the most convenient 
way to calculate the exponent $A$ is by the use of a classical trajectory $x(\tau)$, $y(\tau)$ in imaginary time $t=i\tau$ 
[\cite{COLEMAN1,COLEMAN2,MILLER,SCHMID1,SCHMID2}]. The trajectory goes in a classically forbidden area (under the barrier)
and connects two classically allowed regions. The classical action, constructed by means of this trajectory, is called 
Euclidean action and determines WKB-type exponent $A$. The method of classical trajectories is relatively simple since it 
allows to determine the tunneling probability in the main (no pre-exponent) approximation $\exp(-A)$ just only solving 
Newton's equation of motion. 

The problem of quantum tunneling in a magnetic field was addressed in Refs.[\cite{SHKL1,SHKL2,THOUL}]. In the Landau gauge 
there is a parabolic gauge potential $m\omega^{2}_{c}(x-x_{0})^{2}/2$ superimposed upon the tunnel barrier potential. The 
cyclotron frequency is $\omega_{c}=|e|H/mc$ and tunneling occurs in the $x$-direction. If the tunnel barrier is not a 
constant, containing weak impurity centers, $x_{0}$ becomes spatially dependent resulting in a variable gauge potential
of a sawtooth shape (parabolic segments) instead of pure parabolic one. This potential is ``pinned'' by impurities, 
separated by the characteristic distance $b$, and repeats their positions [\cite{SHKL1,SHKL2}]. In the regime of a strong 
magnetic field, when the energy $m\omega^{2}_{c}b^{2}/2$ exceeds a height of the tunnel barrier, an electron tunnels 
incoherently through each peak of the gauge potential. For this incoherent motion the total probability of tunneling is a 
product of partial ones.

This process can be elegantly described in terms of classical trajectories in imaginary time [\cite{BLATT,GOROKH}]. In 
this formalism the coordinate $x(\tau)$ remains real but $y(\tau)=-i\eta(\tau)$ becomes imaginary. For the strong magnetic
field a kinetic energy is not important and one can consider a massless limit. 

When the magnetic field is not strong, kinetic energy enters the game and a scenario of tunneling can be dramatically 
different. Below we consider a static two-dimensional tunnel potential which, in the barrier region, is only 
$y$-dependent, tunneling occurs in the $x$-direction, and the static magnetic field is directed along the $z$-axis. As in 
Refs.~[\cite{BLATT,GOROKH}], we use a method of classical trajectories to find a tunneling probability. 

As shown in the paper, when the magnetic field is close to a certain value $H_{R}$ the probability of tunneling 
$w\sim\exp(-A)$ through a long barrier becomes not exponentially small since $A\rightarrow 0$. This process can be called 
long distance tunneling. It strongly contrasts to WKB-like tunneling which is associated solely with an exponential decay 
inside a barrier. 

In general, a phenomenon, when $A\rightarrow 0$ at some value of a parameter, is referred to as Euclidean resonance. It was
initially studied in papers [\cite{IVLEV1,IVLEV2,IVLEV3,IVLEV4,IVLEV5}] for tunneling through nonstationary barriers when 
$A\rightarrow 0$ at a certain value of an ac amplitude. In our case of a static potential barrier in a static magnetic 
field Euclidean resonance also takes place. It occurs at $H=H_{R}$ and it is a reason of the above mentioned long distance
tunneling. 

A schematic interpretation of long distance tunneling is associated with formation of a variable gauge potential, as for a
barrier with impurities [\cite{SHKL1,SHKL2}]. In our case there are no impurities and the variable gauge potential is 
formed due to an intrinsic mechanism. It is periodic in the direction of tunneling. An electron can move in this periodic 
potential in a way similar to conventional motion in a periodic structure with a narrow energy band. In this structure 
tunneling processes through subsequent periodic barriers are strongly coherent and there is no exponential decay of a 
moving wave packet [\cite{ZIMAN}]. One can say that the magnetic field sets a long distance under-barrier coherence. 

The goal of this paper is not to investigate long distance tunneling in details but rather to establish the phenomenon. One
can conclude that the nature allows, in principle, a long distance motion under an almost classical potential barrier 
which strongly contrasts to WKB-like tunneling. 
\section{TUNNELING IN ONE DIMENSION}
\label{WKB}
\begin{figure}
\includegraphics[width=7cm]{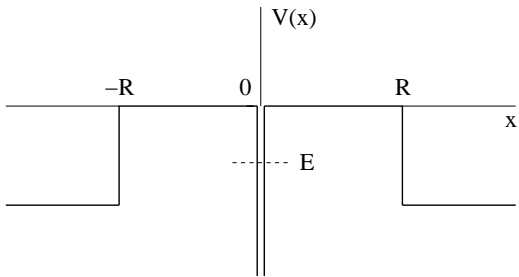}
\caption{\label{fig1}An one-dimensional symmetric potential. The well in the center is of a $\delta$-type. An electron 
from the quasi-discrete state with the energy $E$ can tunnel through the right (left) barrier.}
\end{figure}

Suppose that an one-dimensional potential $V(x)$, shown in Fig.~\ref{fig1}, is symmetric, $V(-x)=V(x)$ and $V(x)=0$ at 
$0<x<R$. At $x=R$ the potential has a jump as for a rectangular barrier. Close to the point
$x=0$ the potential has the form 
\begin{equation} 
\label{1}
V(x)=-\hbar\sqrt{\frac{2|E|}{m}}\hspace{0.1cm}\delta(x)
\end{equation}
where $E<0$ is a discrete energy level in the $\delta$-potential if the barrier would be infinitely long 
($R\rightarrow\infty$). Since the barrier length $R$ is big but finite the energy level turns into a quasi-level with an 
exponentially small width. The probability of tunneling from the localized state at $x=0$ through the barrier to the right
(or left) is given by the WKB expression [\cite{LANDAU}]
\begin{equation} 
\label{2}
w\sim\exp(-A_{WKB});\hspace{0.5cm}A_{WKB}=\frac{2\sqrt{2m|E|}}{\hbar}R
\end{equation}
The result (\ref{1}) is valid with the exponential accuracy (no pre-exponential factor).
\section{TUNNELING IN A MAGNETIC FIELD}
\label{magn}
In this Section we consider an electron localized in the plane $\{x,y\}$. A magnetic field $H$ is directed along the 
$z$-axis. 
\subsection{Quantum mechanical problem}
Suppose the motion of an electron in the $\{x,y\}$ plane to occur in the potential $u(y)+V(x)$ where $V(x)$ is plotted in
Fig.~\ref{fig1} (with the property (\ref{1})) and $u(y)$ is even with respect to $y$. Below we consider the form
\begin{equation} 
\label{3}
u(y)=u_{0}\left(\frac{y^{2}}{a^{2}}+\frac{y^{4}}{a^{4}}\right)
\end{equation}
with a positive $u_{0}$. Then the Schr\"{o}dinger equation with the vector potential $\vec A=\{-Hy,0,0\}$ has the form 
[\cite{LANDAU}]
\begin{eqnarray} 
\label{4}
&&-\frac{\hbar^{2}}{2m}\left(\frac{\partial}{\partial x}-\frac{im\omega_{c}}{\hbar}y\right)^{2}\psi -
\frac{\hbar^{2}}{2m}\frac{\partial^{2}\psi}{\partial y^{2}}+\left[u(y)+V(x)\right]\psi
\nonumber\\
&&=E\psi
\end{eqnarray}
Strictly speaking, an exact Schr\"{o}dinger equation for the potential in Fig.~\ref{fig1} is not static since there is a 
weak leakage through the potential barrier. But the dynamical corrections are exponentially small (generic with the 
expression (\ref{2})) and the static equation (\ref{4}) is valid with this accuracy. If to take a symmetric $V(x)$ (with 
the property (\ref{1})) which is zero at $0<x$ excepting the $\delta$-well of the type (\ref{1}) at $x=R$, than the static 
Schr\"{o}dinger equation (\ref{4}) would be exact with any accuracy. 

Below we consider a wave function of the form
\begin{figure}
\includegraphics[width=7cm]{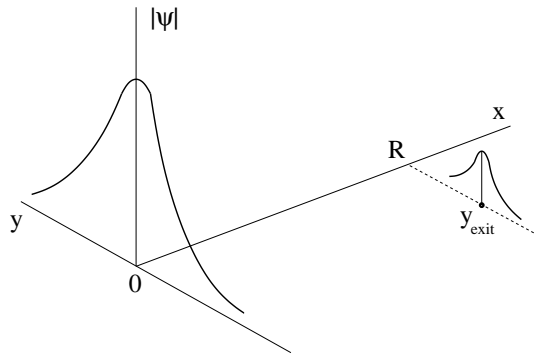}
\caption{\label{fig2}The wave function at $x=0$ (before tunneling) and $x=R$ (after tunneling) in the two-dimensional case 
with the magnetic field.}
\end{figure}
\begin{equation} 
\label{5}
\psi(x,y)=\sum^{\infty}_{n=0}\left[P_{n}(x)\varphi_{2n}(y)+iQ_{n}(x)\varphi_{2n+1}(y)\right]
\end{equation}
where the real functions $P_{n}(x)$ and $Q_{n}(x)$ satisfy the relations $P_{n}(-x)=P_{n}(x)$ and $Q_{n}(-x)=-Q_{n}(x)$.
The full set of real eigenfunctions $\varphi_{n}(y)$ is determined by the equation for the harmonic oscillator
\begin{equation} 
\label{6}
-\frac{\hbar^{2}}{2m}\hspace{0.1cm}\frac{\partial^{2}\varphi_{n}}{\partial y^{2}}+\frac{u_{0}}{a^{2}}y^{2}\varphi 
=\frac{\hbar}{a}\sqrt{\frac{u_{0}}{2m}}(1+2n)\varphi 
\end{equation}
Note, that $\varphi_{2n}(y)$ is even with respect to $y$ and $\varphi_{2n+1}(y)$ is odd [\cite{LANDAU}]. It is easy to 
check that a solution of the type (\ref{5}) does not contradict to the equation (\ref{4}). 

It follows from Eqs.~(\ref{5}) and (\ref{1}) that 
\begin{equation} 
\label{7}
\frac{\partial\psi(x,0)}{\partial x}\Bigg |_{x=0}=-\frac{\sqrt{2m|E|}}{\hbar}\psi(0,0),\hspace{0.3cm} 
\frac{\partial\psi(0,y)}{\partial y}\Bigg |_{y=0}=0
\end{equation}
A profile of the wave function is shown in Fig.~\ref{fig2} at $x=0$ and $x=R$. At $x=0$ it corresponds to the conditions
(\ref{7}). On the exit line $x=R$ a modulus of the wave function reaches a maximum at a certain point $y_{\rm{exit}}$ which
is determined by a solution of the Schr\"{o}dinger equation. Therefore, a probability of tunneling can be defined as
\begin{equation} 
\label{8}
w=\Bigg |\frac{\psi\left(R,y_{\rm{exit}}\right)}{\psi\left(0,0\right)}\Bigg |^{2}
\end{equation}
\subsection{Quasi-classical approach}
Since the potential barrier in Fig.~\ref{fig1} is weakly transparent one can use quasi-classical approximation 
[\cite{LANDAU}] for the wave function under the barrier, in the region $0<x<R$,
\begin{equation} 
\label{9}
\psi(x,y)\sim\exp\left[\frac{i}{\hbar}S(x,y)\right]
\end{equation}
where $S(x,y)$ is a classical action satisfying the equation of Hamilton-Jacobi
\begin{figure}
\includegraphics[width=7cm]{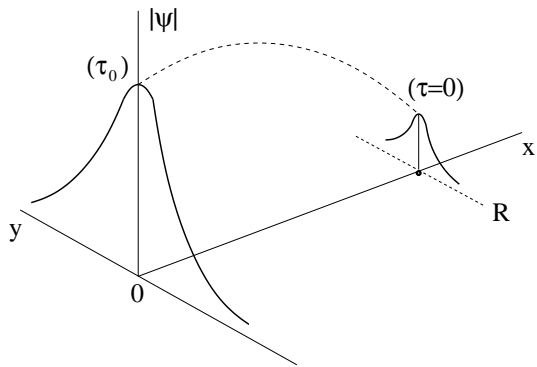}
\caption{\label{fig3}The wave function $\psi(x,y)$ at $x=0$ and $x=R$ when $y_{\rm exit}=0$. The dashed curve shows 
schematically a ``bypass'' provided by the complex classical trajectory operating with $\psi(x,-i\eta)$.}
\end{figure}
\begin{equation} 
\label{10}
\frac{1}{2m}\left(\frac{\partial S}{\partial x}-m\omega_{c}y\right)^{2}+
\frac{1}{2m}\left(\frac{\partial S}{\partial y}\right)^{2}+u(y)=E
\end{equation}
According to Eq.~(\ref{7}), the action $S(x,y)$ obeys the conditions
\begin{equation} 
\label{11}
\frac{\partial S(x,0)}{\partial x}\Bigg |_{x=0}=i\sqrt{2m|E|}\hspace{0.1cm},\hspace{0.5cm}
\frac{\partial S(0,y)}{\partial y}\Bigg |_{y=0}=0
\end{equation}
The equation (\ref{9}) defines the wave function only with the main (exponential) accuracy. This means that a 
pre-exponential factor, which is less significant in quasiclassical case, is neglected. As follows from Eqs.~(\ref{8}) and
(\ref{9}), the tunneling probability in the main approximation has the form 
\begin{equation} 
\label{12}
w\sim\exp\left\{-\frac{2}{\hbar}{\rm Im}\left[S\left(R,y_{\rm exit}\right)-S\left(0,0\right)\right]\right\}
\end{equation}
\subsection{Classical trajectories}
Let us consider a classical motion of an electron in the potential $u(y)$ 
\begin{equation} 
\label{13}
m\frac{\partial^{2}x}{\partial t^{2}}=-m\omega_{c}\frac{\partial y}{\partial t},\hspace{0.5cm}
m\frac{\partial^{2}y}{\partial t^{2}}=m\omega_{c}\frac{\partial x}{\partial t}-\frac{\partial u(y)}{\partial y}
\end{equation}
with the initial conditions
\begin{equation} 
\label{14}
x(t_{0})=y(t_{0})=0,\hspace{0.3cm}\frac{\partial x(t)}{\partial t}\Bigg |_{t_{0}}=i\sqrt{\frac{2|E|}{m}}\hspace{0.1cm},
\hspace{0.3cm}\frac{\partial y(t)}{\partial t}\Bigg |_{t_{0}}=0
\end{equation}
The equations of motion (\ref{13}) together with the conditions (\ref{14}) determine a classical trajectory fully. 

The general expression
\begin{equation} 
\label{15}
dS(x,y)=\frac{\partial S}{\partial x}dx+\frac{\partial S}{\partial y}dy
\end{equation}
along the classical trajectory $\{x(t),y(t)\}$ turns into
\begin{equation} 
\label{16}
dS=\left(\frac{\partial S}{\partial x}\hspace{0.1cm}\frac{\partial x}{\partial t}+
\frac{\partial S}{\partial y}\hspace{0.1cm}\frac{\partial y}{\partial t}\right)dt
\end{equation}
Along the trajectory classical momenta $p_{x}$ and $p_{y}$ obey the conditions
\begin{equation} 
\label{17}
p_{x}=\frac{\partial S}{\partial x}=m\frac{\partial x}{\partial t}+m\omega_{c}y,\hspace{0.5cm}
p_{y}=\frac{\partial S}{\partial y}=m\frac{\partial y}{\partial t}
\end{equation}
One can consider the action $S(x,y)$ not in the full two-dimensional plane but only on the classical trajectory 
$S\left[x(t),y(t)\right]$. The initial conditions for the classical trajectory (\ref{14}) are chosen in order to match the
conditions (\ref{11}) for the action. Therefore, as follows from Eq.~(\ref{16}),
\begin{equation} 
\label{18}
S\left[x(t),y(t)\right]-S\left(0,0\right)=
\int^{t}_{t_{0}}dt_{1}\left(p_{x}\frac{\partial x}{\partial t_{1}}+p_{y}\frac{\partial y}{\partial t_{1}}\right)
\end{equation}
According to classical mechanics, 
\begin{equation} 
\label{19}
p_{x}\frac{\partial x}{\partial t}+p_{y}\frac{\partial y}{\partial t}=L+H
\end{equation}
where $L$ is a Lagrangian and $H$ is a Hamiltonian. Using the expression for a Lagrangian in the magnetic field 
[\cite{LANDAU1}] and taking $H=E$, one can rewrite Eq.~(\ref{18}) in the form
\begin{eqnarray} 
\label{20}
S\left[x(t),y(t)\right]-S\left(0,0\right)=
\int^{t}_{t_{0}}dt_{1}\bigg[\frac{m}{2}\left(\frac{\partial x}{\partial t_{1}}\right)^{2}\\
\nonumber
+\frac{m}{2}\left(\frac{\partial y}{\partial t_{1}}\right)^{2}+m\omega_{c}y\frac{\partial x}{\partial t_{1}}-u(y)+E\bigg]
\end{eqnarray}

The total energy of an electron, which moves along the classical trajectory in the magnetic field, is
\begin{equation} 
\label{21}
E=\frac{m}{2}\left(\frac{\partial x}{\partial t}\right)^{2}
+\frac{m}{2}\left(\frac{\partial y}{\partial t}\right)^{2}+u\left(y\right)
\end{equation}
\subsection{Complex classical trajectories}
Our goal is to apply Eq.~(\ref{20}) for calculation of the tunneling probability (\ref{12}). In other words, one should 
find a trajectory which connects two top points in Fig.~\ref{fig2}. A problem is that there is no conventional classical 
trajectory under the barrier since at $E<u(y)$ the kinetic part of the total energy (\ref{21}) becomes negative. 
Nevertheless, there is a famous method to avoid this difficulty. One should formally use imaginary time $t=i\tau$ 
[\cite{COLEMAN1,COLEMAN2,MILLER,SCHMID1,SCHMID2}]. In this case the kinetic part of the total energy (\ref{21}) is negative
which allows a motion under the barrier. One should put $t=i\tau$ in all previous equations for the classical trajectories.

As follows from the classical equations of motion (\ref{13}), the coordinate $x(\tau)$ remains real in imaginary time but 
the other coordinate becomes imaginary $y(\tau)=-i\eta(\tau)$. Generally speaking, this results in an imaginary terminal
point $y(t)$ in the left-hand side of Eq.~(\ref{20}). It means that the trajectory leads to an unphysical terminal point
which does not make the method applicable. However, the method becomes physically correct if, under certain conditions, 
the terminal point turns to zero since there is no difference how to approach zero, from real or imaginary direction. 

As shown below, the exit point $y_{\rm exit}$ in Fig.~\ref{fig2} turns to zero at certain values of the magnetic field. 
This situation relates to Fig.~\ref{fig3}. In this case there is a trajectory connecting the top points in Fig.~\ref{fig3}
drawn by the dashed curve. For this trajectory it is convenient to choose in Eq.~(\ref{20}) $t=0$ and $t_{0}=i\tau_{0}$ 
where $\tau_{0}$ is some ``moment'' of imaginary time to be determined from the condition $x(0)=R$. In this situation the 
left-hand side of Eq.~(\ref{20}) coincides with the exponential in Eq.~(\ref{12}) which justifies the method of 
trajectories.

Now one can formulate a problem of calculation of tunneling probability in terms of complex classical trajectories. Again,
this method, in its present formulation, works solely at the certain values of the magnetic field when $y_{\rm exit}$ is 
zero. The probability of tunneling 
\begin{equation} 
\label{22}
w\sim\exp\left(-A\right)
\end{equation}
according to Eqs.~(\ref{12}) and (\ref{20}), is expressed through the Euclidean action
\begin{eqnarray} 
\label{23}
A&=&\frac{2}{\hbar}\int^{\tau_{0}}_{0}\bigg\{\frac{m}{2}\left(\frac{\partial x}{\partial\tau}\right)^{2}-
\frac{m}{2}\left(\frac{\partial\eta}{\partial\tau}\right)^{2}+m\omega_{c}\eta\hspace{0.1cm}\frac{\partial x}{\partial\tau}
\nonumber\\
&&{}+u\left(-i\eta\right)-E\bigg\}
\end{eqnarray}
To obtain the expression (\ref{23}) one should put $t_{1}=i\tau$, $t=0$, $t_{0}=i\tau_{0}$, and $y=-i\eta$ in 
Eq.~(\ref{20}). The coordinates $x(\tau)$ and $\eta(\tau)$ in Eq.~(\ref{23}) are solutions of the equation of motion 
following from Eq.~(\ref{13})
\begin{equation} 
\label{24}
m\frac{\partial^{2}x}{\partial\tau^{2}}=-m\omega_{c}\frac{\partial\eta}{\partial\tau},\hspace{0.5cm}
m\frac{\partial^{2}\eta}{\partial\tau^{2}}=
-m\omega_{c}\frac{\partial x}{\partial\tau}-\frac{\partial u(-i\eta)}{\partial\eta}
\end{equation}
The conditions to the equations of motion (\ref{24}) are
\begin{equation} 
\label{25}
\eta(0)=\eta(\tau_{0})=0,\hspace{0.3cm}\frac{\partial\eta}{\partial\tau}\bigg |_{0}=
\frac{\partial\eta}{\partial\tau}\bigg |_{\tau_{0}}=0
\end{equation}
and 
\begin{equation} 
\label{26}
x(0)=R,\hspace{0.3cm}x(\tau_{0})=0,\hspace{0.3cm}\frac{\partial x}{\partial\tau}\bigg |_{0}=
\frac{\partial x}{\partial\tau}\bigg |_{\tau_{0}}=-v
\end{equation}
where $v=\sqrt{2|E|/m}$. The total energy follows from Eq.~(\ref{21})
\begin{equation} 
\label{27}
E=-\frac{m}{2}\left(\frac{\partial x}{\partial\tau}\right)^{2}
+\frac{m}{2}\left(\frac{\partial\eta}{\partial\tau}\right)^{2}+u\left(-i\eta\right)
\end{equation}

A solution of Eqs.~(\ref{24}) depends on four independent constants. There are two types of solution. One of them 
corresponds to even $\partial x(\tau)/\partial\tau$ and $\eta(\tau)$ with respect to $(\tau -\tau_{0}/2)$. The second 
solution relates to odd ones. One should choose the first solution. This defines one of the four constants. The type of 
solution chosen depends on three constants and should satisfy the four conditions (\ref{25}) and (\ref{26}) taken at 
$\tau =0$. This is possible since there is an additional free parameter $\tau_{0}$. 

The method of complex classical trajectories allows to connect the two physical points in Fig.~\ref{fig3} which 
characterize states before ($\tau=\tau_{0}$) and after ($\tau=0$) tunneling. This connection is shown schematically by the 
dashed curve in Fig.~\ref{fig3}. One can say that this curve provides, in the action $S(x,y)$, a ``bypass'' of the region
$0<x<R$ through the complex plane operating with the function $S(x,-i\eta)$. Inside that region the wave function can be 
very complicated and an advantage of the trajectory method is that this complicated behavior remains in shadow since we
are interested by only terminal points. 
\section{LONG DISTANCE TUNNELING}
\label{long}
In this Section we give a further consideration of tunneling in the magnetic field by solving classical equations of motion
to obtain particular results.
\subsection{Classical trajectory} 
A solution of the first equation (\ref{24}) has the form
\begin{equation} 
\label{28}
\frac{\partial x}{\partial\tau}=-(v+\omega_{c}\eta)
\end{equation}
With Eq.~(\ref{28}) the expression (\ref{27}) turns to 
\begin{equation} 
\label{29}
\frac{m}{2}\left(\frac{\partial\eta}{\partial\tau}\right)^{2}+v(\eta)=E
\end{equation}
where the potential energy is
\begin{equation} 
\label{30}
v(\eta)=u(-i\eta)-\frac{m}{2}\left(v+\omega_{c}\eta\right)^{2}
\end{equation}
As follows from (\ref{29}), a dynamics of the transverse component $\eta$ of the classical trajectory is analogous to a 
classical motion in the potential (\ref{30}).

In the equation (\ref{30}) $v(0)=-mv^{2}/2=E$. If $\eta(0)=0$ then, as follows from (\ref{29}), 
$\partial\eta/\partial\tau$ is also zero at $\tau=0$. Therefore, to get the same conditions at $\tau=\tau_{0}$ a motion in
the potential (\ref{30}) should be periodic and the potential (\ref{30}) should have a form of a potential well. 
\begin{figure}
\includegraphics[width=5cm]{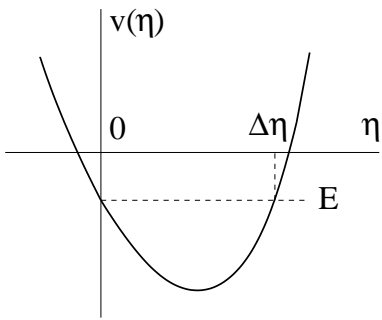}
\caption{\label{fig4}The effective potential for a transverse motion in imaginary time forms a well.}
\end{figure}
In our case, the potential (\ref{30}) has the form
\begin{equation} 
\label{31}
v(\eta)=u_{0}\left(\frac{\eta^{4}}{a^{4}}-\frac{\eta^{2}}{a^{2}}\right)-\frac{m}{2}\left(v+\omega_{c}\eta\right)^{2}
\end{equation}
The potential (\ref{31}), with the parameters chosen in Sec.~\ref{calc}, is plotted in Fig.~\ref{fig4} where 
$v(\Delta\eta)=E$. 
\subsection{Choice of a potential $u(y)$}
One should make a remark about the shape of the potential in Fig.~\ref{fig4}. This shape substantially depends on 
analytical properties of the function (\ref{3}) in the complex plane. If to take in Eq.~(\ref{3}) a pure quadratic 
potential $u(y)=u_{0}y^{2}/a^{2}$ or a pure harmonic one $u(y)=u_{0}\left(1-\cos y/a\right)$ then it would be no well in 
$v(\eta)$ and the considered method does not work. The potential (\ref{3}) is not unique one resulting in the well. For 
example, the form $u(y)=u_{0}\left(1-\cos y/a\right)^{2}$ is also suitable. We do not completely analyze a connection of a
shape of $v(\eta)$ with analytical properties of $u(y)$ and only restrict ourself by the simplest form (\ref{3}). 
\subsection{Periodic motion}
\begin{figure}
\includegraphics[width=6cm]{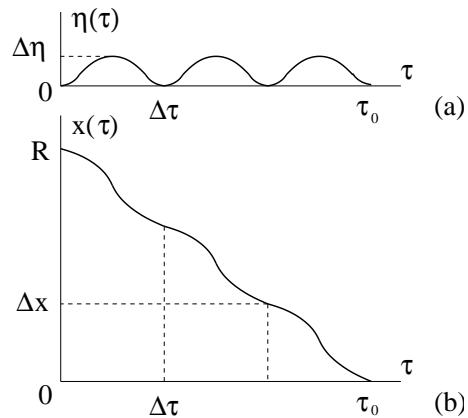}
\caption{\label{fig5}The classical trajectory in imaginary time. It is chosen $N=3$. (a) The transverse component 
($y=-i\eta$). (b) The motion in the direction of tunneling.}
\end{figure}

The periodic motion of $\eta(\tau)$ is shown in Fig.~\ref{fig5}(a). The function $x(\tau)$, determined by (\ref{28}), is 
drawn in Fig.~\ref{fig5}(b). A period $\Delta\tau$ of oscillations, according to Eq.~(\ref{29}), is
\begin{equation} 
\label{32}
\Delta\tau =\sqrt{2m}\int^{\Delta\eta}_{0}\frac{d\eta}{\sqrt{E-v(\eta)}}
\end{equation}
Each cycle of $\eta(\tau)$ in Fig.~\ref{fig5}(a) results in the translation of $x(\tau)$ by $\Delta x$ determined by
\begin{equation} 
\label{33}
\Delta x=\sqrt{2m}\int^{\Delta\eta}_{0}d\eta\frac{v+\omega_{c}\eta}{\sqrt{E-v(\eta)}}
\end{equation}
\subsection{Euclidean action}
The probability of tunneling (\ref{22}) is determined by the Euclidean action (\ref{23}). By means of Eq.~(\ref{28}) the 
action (\ref{23}) reads
\begin{equation} 
\label{34}
A=\frac{2}{\hbar}\int^{\tau_{0}}_{0}d\tau\left[-\frac{m}{2}\left(\frac{\partial\eta}{\partial\tau}\right)^{2}
+v(\eta)-E-mv\frac{\partial x}{\partial\tau}\right]
\end{equation}
With the expression (\ref{29}) the action (\ref{34}) takes the form
\begin{equation} 
\label{35}
A=A_{WKB}-\frac{2m}{\hbar}\int^{\tau_{0}}_{0}d\tau\left(\frac{\partial\eta}{\partial\tau}\right)^{2}
\end{equation}
where $A_{WKB}$ is determined by Eq.~(\ref{2}) and comes from the last term in Eq.~(\ref{34}). $A_{WKB}$ in Eq.~(\ref{35})
is connected with the motion in the direction joining the two wells and it is generic with the conventional under-barrier 
action in a multi-dimensional case [\cite{COLEMAN1,COLEMAN2,SCHMID1,SCHMID2}]. The second term in Eq.~(\ref{35}) is a
counter-part originated from the transverse kinetic energy. It is negative since for a classical trajectory in a magnetic 
field $y^{2}=-{\eta}^{2}$ is negative.  

As mentioned above, the method of trajectories is applicable when the exit point $y_{\rm exit}$ in Fig.~\ref{fig2} is zero.
This is possible when the barrier length $R$ coincides with an integer number of periods $\Delta x$ ($R=N\Delta x$ and 
also $\tau_{0}=N\Delta\tau$), since at the end of each period the condition $\eta =0$ holds. Fig.~\ref{fig5} is plotted 
for $N=3$. By means of (\ref{29}), the last term in Eq.~(\ref{35}) can be written as $(-N\Delta A)$ where 
\begin{equation} 
\label{36}
\Delta A=\frac{4\sqrt{2m}}{\hbar}\int^{\Delta\eta}_{0}d\eta\sqrt{E-v(\eta)}
\end{equation}
The action (\ref{35}) is linear with respect to $R$ and takes the form
\begin{equation} 
\label{37}
A=A_{WKB}-\Delta A\frac{R}{\Delta x}=\left(\frac{2\sqrt{2m|E|}}{\hbar}-\frac{\Delta A}{\Delta x}\right)R
\end{equation}
The second terms in Eq.~(\ref{37}) are counter-parts resulting from the transverse kinetic energy.

The condition of applicability of the presented method of trajectories
\begin{equation} 
\label{38}
R=N\Delta x(H)
\end{equation}
imposes a restriction on possible values of the magnetic field $H=h_{N}$. 
\subsection{Euclidean resonance}
As one can see from Eq.~(\ref{37}), the action $A$ is reduced compared to its WKB part $A_{WKB}$. Under variation of the 
magnetic field, as shown below, one can encounter a situation when $A$ becomes small and formally tends to zero, at a 
certain value $H_{R}$ of the magnetic field, as $A\sim (H_{R}-H)R$. This means that the tunneling probability becomes not 
exponentially small at $H\rightarrow H_{R}$ even for a long barrier (big $R$) [\cite{IVLEV6}]. The quantity $H_{R}$ plays 
a role of a ``resonance'' magnetic field. Since the probability to cross a long barrier is not small this process can be 
called long distance tunneling. 

The phenomenon when $A\rightarrow 0$ at some value of a parameter is referred to as Euclidean resonance. Euclidean 
resonance was established initially for tunneling through a nonstationary barrier when the action turned to zero at a
certain ac amplitude [\cite{IVLEV1,IVLEV2,IVLEV3,IVLEV4,IVLEV5}]. Euclidean resonance occurs also in our case of a static 
barrier in a static magnetic field giving rise to long distance tunneling. 

The method of trajectories used corresponds to one-instanton approach when a probability is determined by a small
$\exp(-A)$ as in Eq.~(\ref{22}). A trajectory in imaginary time is called instanton. Indeed, at $H<H_{R}$ the probability 
Eq.~(\ref{22}) is small. When $H$ is close to $H_{R}$ the quantity $\exp(-A)$ is not small and one should apply a 
multi-instanton approach which accounts all powers of the exponent (\ref{22}). This approach is beyond our one-instanton 
method and, strictly speaking, we do not know $w(H)$ at $H>H_{R}$. 

\begin{figure}
\includegraphics[width=6cm]{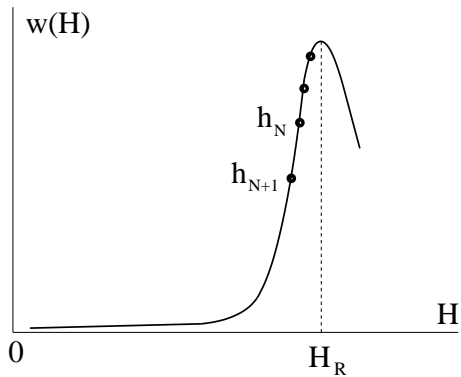}
\caption{\label{fig6}The probability of tunneling as a function of a magnetic field has the peak at $H_{R}$ (Euclidean 
resonance). The values of the magnetic field $h_{N}$ correspond to the total distance $R=N\Delta x$.}
\end{figure}

Nevertheless, one has to expect that at $H>H_{R}$ the tunneling probability also decreases. An indication of that is a 
small probability found in Refs.[\cite{SHKL1,SHKL2,BLATT,GOROKH}] at high magnetic field, $|E|\ll m\omega^{2}_{c}a^{2}$, 
which is bigger than $H_{R}$. Therefore, one should expect a peak of $w(H)$ at $H=H_{R}$ as drawn in Fig.~\ref{fig6}. As 
shown in Sec.~\ref{calc}, the width of the left-hand part of
the peak can be estimated as $H_{R}/A_{WKB}$. The dots on the curve in Fig.~\ref{fig6} correspond to applicability of the 
present method of trajectories. In other words, we know the probability of tunneling only at discrete points on the curve 
in Fig.~\ref{fig6}. The full smooth curve goes through these discrete points. If, occasionally, $H_{R}$ is close to one of
$h_{N}$, then one can approach the top of the curve in Fig.~\ref{fig6} within the method used. 
\section{CALCULATION OF TUNNELING PROBABILITY}
\label{calc}
In this Section we perform a calculation of tunneling probability on the basis of the formalism developed in 
Sec.~\ref{long}. For simplicity we choose $u_{0}=|E|$ in Eq.~(\ref{31}). Then the action (\ref{37}) takes the form 
\begin{equation}
\label{39}
A=\left[1-\frac{f_{1}(p)}{f_{2}(p)}\right]A_{WKB}
\end{equation}
where
\begin{equation}
\label{40}
p=\frac{m\omega^{2}_{c}a^{2}}{2|E|}
\end{equation}
The functions are defined by the relations 
\begin{equation}
\label{41}
f_{1}(p)=2\int^{z_{0}}_{0}dz\sqrt{z^{2}-z^{4}+2z\sqrt{p}+pz^{2}}
\end{equation}
and
\begin{equation}
\label{42}
f_{2}(p)=2\int^{z_{0}}_{0}dz\frac{1+z\sqrt{p}}{\sqrt{z^{2}-z^{4}+2z\sqrt{p}+pz^{2}}}
\end{equation}
where $z=\eta/a$ and the limit of integration $z_{0}=\Delta\eta/a$ corresponds to a zero of the square root. With these 
definitions,  
\begin{equation}
\label{43}
\Delta A=\frac{2a\sqrt{2m|E|}}{\hbar}f_{1}(p),\hspace{0.5cm}\Delta x=af_{2}(p)
\end{equation}
Eq.~(\ref{40}) leads to the relation
\begin{equation}
\label{44}
H\simeq 3.37\sqrt{p}\hspace{0.1cm}\frac{\sqrt{|E|({\rm eV})}}{a(\AA)}\times 10^{4}({\rm Tesla})
\end{equation}
When $p$ is close to the value $p_{R}\simeq 1.76$, as follows from numerical calculations, 
$f_{1}(p_{R})=f_{2}(p_{R})\simeq 6.52$ and the action (\ref{39}) reads
\begin{equation}
\label{45}
A\simeq 0.31\left(p_{R}-p\right)A_{WKB}
\end{equation}
According to Eq.~(\ref{44}), $p_{R}$ determines the resonance magnetic field $H_{R}$
\begin{equation}
\label{46}
H_{R}\simeq 4.47\hspace{0.1cm}\frac{\sqrt{|E|({\rm eV})}}{a(\AA)}\times 10^{4}({\rm Tesla})
\end{equation}
As one can see from Eq.~(\ref{40}), $H_{R}$ corresponds to the condition $m\omega^{2}_{c}a^{2}\sim |E|$. When $H$ is close
to $H_{R}$
\begin{equation}
\label{47}
A\simeq 1.1\frac{H_{R}-H}{H_{R}}A_{WKB}
\end{equation}
Within the approach used, the magnetic field $H$ has to provide the condition $\exp(-A)\ll 1$. This means that $H$ should 
be less than $H_{R}$ and not very close to this value. The probability of tunneling (\ref{22}) with the expression 
(\ref{47}) describes the left-hand part of the curve in Fig.~\ref{fig6}. Discrete points $h_{N}$ on that curve correspond 
to values $p_{N}$ in Eq.~(\ref{44}) which are determined by the condition following from (\ref{38}) and (\ref{43})
\begin{equation}
\label{48}
\frac{R}{a}=Nf_{2}\left(p_{N}\right)
\end{equation}
Close to $p_{R}$ one can use the expansion $f_{2}(p)\simeq f_{2}(p_{R})-0.33(p_{R}-p)$. Together with the definition 
(\ref{40}) one can obtain positions of the dots in Fig.~\ref{fig6}
\begin{equation}
\label{49}
\frac{h_{N}}{H_{R}}=1-0.43\left(6.52-\frac{R}{Na}\right)
\end{equation}
The discrete values on the left-hand part of the curve in Fig.~\ref{fig6} is described by the formula
\begin{equation}
\label{50}
w(h_{N})\sim\exp\left(-1.1\hspace{0.1cm}\frac{H_{R}-h_{N}}{H_{R}}A_{WKB}\right)
\end{equation}
which follows from Eqs.~(\ref{22}) and (\ref{47}). Eqs.~(\ref{49}) and (\ref{50}) are valid when $h_{N}$ is less than 
$H_{R}$ and is close to this value. 

The potential barrier, chosen in Sec.~\ref{magn}, rather corresponds to tunneling between quantum wires. A barrier between
wires should be one of a variety of analytical forms resulting in a well of the potential $v(\eta)$. In a real experiment,
when an inter-wire potential is created by some method, it is hard to get a certain analytical form of $u(y)$.
Nevertheless, among various realizations of $u(y)$ a proper analytical form can exist. Let us choose in the potential 
(\ref{3}) $a=140\AA$. The discrete energy level $E=-10^{-3}$eV relates to experimental values [\cite{WIEL}]. With these 
parameters one gets $H_{R}\simeq 10$~(Tesla), $\Delta A\simeq 29.6$, $\Delta x\simeq 913~\AA$, and 
$A_{WKB}\simeq 0.032R~(\AA)$. 

It is informative to note that for the barrier length $R=\Delta x$ the tunneling probability without a magnetic field 
(\ref{2}) is $10^{-13}$ and for $R=3\Delta x$ it is $10^{-39}$. The magnetic field may turn these probabilities into not
small values. 
\section{INTERPRETATION OF LONG DISTANCE TUNNELING}
\label{interpr}
In this Section we propose a scheme for explanation of long distance tunneling in physical terms. 

The method of trajectories provides not only a wave function $\psi(N\Delta x,0)$ but also 
\begin{equation}
\label{51}
\big |\psi(n\Delta x,0)\big |^{2}\sim
\exp\left[-\left(\frac{2\sqrt{2m|E|}}{\hbar}-\frac{\Delta A}{\Delta x}\right)n\Delta x\right]
\end{equation}
with $n=1,2,...N$. It is possible since the trajectory in Fig.~\ref{fig5} passes through the physical points 
$\{n\Delta x,0\}$. The expression (\ref{51}) follows from Eq.~(\ref{37}) if to substitute 
$R=N\Delta x\rightarrow n\Delta x$. Close to $H_{R}$ there is no exponential decay of $\psi(n\Delta x,0)$ with respect to
$n$. This says for an almost periodic (at least within the exponential accuracy) character of the wave function along the
$x$-axis. 

A free electron in a magnetic field moves in the Landau gauge potential $m\omega^{2}_{c}(x-x_{0})/2$ [\cite{LANDAU}]. In 
presence of a tunnel barrier the motion of an electron can be interpreted in terms of formation of a variable gauge
potential when $x_{0}$ becomes a function of coordinates. For example, this interpretation is reasonable when there are 
impurities inside a tunnel barrier which ``pin'' $x_{0}$ [\cite{SHKL1,SHKL2}]. 

In our case of an almost periodic wave function one can also propose a schematic interpretation in terms of a variable 
gauge potential centered around the points $x_{0}=n\Delta x$. A particle moves in this periodic potential in a way similar
to conventional motion in a periodic structure with a narrow energy band where tunneling processes through subsequent 
periodic barriers are strongly coherent. For a conventional periodic structure a wave packet can pass over a long distance
with no exponential reduction in amplitude. This is analogous to our tunneling when the magnetic field is close to $H_{R}$
and decay of a packet with distance is weak. One can say that the magnetic field sets an under-barrier coherence. 
\section{ROLE OF DISSIPATION}
\label{dissip}
The under-barrier coherence can be influenced by a dissipation. As known, a dissipation results in a finite width 
$\delta E$ of energy levels in a well and also disturbs an under-barrier motion, according to Caldeira and Leggett 
[\cite{LEGGETT}]. In the classical dynamics dissipation corresponds to the form $m\ddot x+m\gamma\dot x$. Using the theory
[\cite{LEGGETT}], one can obtain (we omit details) the criterion $\gamma N\Delta\tau<1$ when dissipation does not 
influence the frictionless motion. Since $\delta E\sim\hbar\gamma$ this criterion is equivalent to 
\begin{equation}
\label{52}
\frac{\delta E}{E}<\frac{1}{A_{WKB}}
\end{equation}
When $\delta E/E\sim 0.01$ [\cite{WIEL}], with the parameters chosen in Sec.\ref{calc}, one can estimate $R<3000\AA$. For 
a bigger $R$ dissipation modifies long distance tunneling but not destroys it. This is a matter of a further research. 
Also one can show that a non-homogeneity $\delta u(x)$ (including applied voltage) of a barrier in the $x$-direction does 
not violate the above results as soon as $\delta u$ is smaller than an energy of the order of $|E|$. 
\section{DISCUSSION}
\label{disc}
A phenomenon when in a tunneling probability $w\sim\exp(-A)$ the quantity $A$ tends to zero at some value of a parameter 
is referred to as Euclidean resonance. A phenomenon of Euclidean resonance is associated with formation of an extended 
quantum coherence under a barrier which substantially differs from a WKB-like decaying wave function. 

This phenomenon was studied initially for tunneling through nonstationary barriers. As follows from the paper, it can also 
occur inside a static potential barrier in a static magnetic field. In this case $A$ tends to zero at the certain magnetic
field $H_{R}$. At a vicinity of $H_{R}$ the tunneling probability rapidly grows up and can reach a value which is not 
exponentially small. This allows tunneling over the distance which is bigger by the factor $H_{R}/(H_{R}-H)$ compared to a
conventional WKB-like one. Euclidean resonance is a reason of long distance tunneling.

In this paper classical trajectories are used to study tunneling through a potential barrier in a magnetic field. This 
wide distributed method is relatively simple and allows to calculate a tunneling probability with the exponential accuracy
(no pre-exponential factor). The method of trajectories works as soon as a classical action exceeds Planck's constant. 
Hence, one can approach the resonance value $H_{R}$ of the magnetic field until $\exp(-A)$ is small. 

The method also enables (by an analytical continuation into complex plane) to select forms of those potential barriers 
which result in long distance tunneling. This reminds a calculation of an integral of a strongly oscillating function when
a deformation of an integration contour toward complex plane leads to a simple evaluation of the integral. In other words, 
a delicate quantum interference can be effectively formulated in terms of complex trajectories. Complex trajectories for 
tunneling in a strong magnetic field ($|E|\ll m\omega^{2}_{c}a^{2}$) were explored in Refs.~[\cite{BLATT,GOROKH}]. 

The method of trajectories used allows to calculate the tunneling probability solely at certain discrete values of the 
magnetic field. A goal of this paper is not to perform a comprehensive investigation of long distance tunneling but just to
establish this phenomenon. For this reason, we consider a mostly convenient potential barrier related to quantum wires. 
One of the next steps in the field is a numerical calculation of the quantum problem. A numerical calculations would 
provide a whole magnetic field dependence of the tunneling probability and a structure of the wave function in two 
dimensions. A numerical calculation constitutes a separate problem. 

An interpretation of long distance tunneling is based on formation of a periodic variable gauge potential under the 
barrier. A particle moves in this periodic potential like in a conventional periodic structure with a narrow energy band 
when subsequent tunneling processes through potential barriers are very coherent. This contrasts with tunneling through a
barrier with impurities [\cite{SHKL1,SHKL2}] where tunneling processes across subsequent barriers (formed by a variable 
gauge potential which is ``pinned'' by impurities) are not coherent. 

Tunneling between quantum wires with physical parameters chosen above can occur if they are separated even by 3000$\AA$.
From the stand point of conventional WKB tunneling such a big separation distance relates to an almost classical potential
barrier with a tunneling probability of $10^{-39}$. The limit 3000$\AA$ is set by dissipation. For a bigger inter-wire 
distance the dissipation affects the coherence but does not destroy it completely. The case of a bigger inter-wire 
distance requires a special consideration. 

Besides particular applications (for example, tunneling between quantum wires), a phenomenon of long distance tunneling 
may be of a more general interest since it strongly contrasts to the known scenario of quantum tunneling. Being based on a
phenomenon of Euclidean resonance, long distance tunneling can be caused by a various types of external perturbations. It
can be an applied magnetic field, nonstationary electric field, and a Coulomb field of charged particles in nuclear 
processes. As preliminary calculations show, a certain static perturbation of a barrier also can result in Euclidean 
resonance and, therefore, in long distance tunneling. It seems to be interesting to study a possibility of a long 
distance coherence in various quantum systems (big size molecules, wide separated artificial quantum wells, etc.). 
\section{CONCLUSIONS}
The nature allows, in principle, a long distance motion under an almost classical static potential barrier which strongly 
contrasts to WKB-like tunneling. An example of this motion, similar to tunneling between quantum wires in a magnetic 
field, is considered in the paper. At moderate dissipation, a particle can tunnel through a long potential barrier as 
through a conventional narrow band periodic structure. The magnetic field sets a long distance coherence under a barrier. 
This phenomenon relates to Euclidean resonance. 

\acknowledgments
I thank A. Barone, A. Bezryadin, G. Blatter, M. Gershenson, V. Geshkenbein, L. Ioffe, J. Knight, G. Pepe, A. Rodriguez, 
A. Ustinov, and R. Webb for discussions of related topics.


\begin{thebibliography}{99}

\bibitem{LANDAU}

L.D. Landau and E.M. Lifshitz, {\it Quantum Mechanics} (Pergamon, New York, 1977).

\bibitem{COLEMAN1}

C.G. Callan and S. Coleman, Phys. Rev. D {\bf 16}, 1762 (1977).

\bibitem{COLEMAN2}

S. Coleman, in {\it Aspects of Symmetry} (Cambridge University Press, Cambridge, 1985).

\bibitem{MILLER}

W.H. Miller, Adv. Chem. Phys. {\bf 25}, 68 (1974).

\bibitem{SCHMID1}

A. Schmid, Ann. Phys. {\bf 170}, 333 (1986).

\bibitem{SCHMID2}

U. Eckern and A. Schmid, in {\it Quantum Tunneling in Condensed Media}, edited by
A. Leggett and Yu. Kagan (North-Holland, Amsterdam, 1992).

\bibitem{SHKL1}

B.I. Shklovskii, Pis'ma Zh. Eksp. Teor. Fiz. {\bf 36}, 43 (1982) [Sov. Phys. JETP Lett. {\bf 36}, 51 (1982)].

\bibitem{SHKL2}

B.I. Shklovskii and A. Efros, Zh. Eksp. Teor. Fiz. {\bf 84}, 811 (1983) [Sov. Phys. JETP {\bf 57}, 470 (1983)].

\bibitem{THOUL}

Q. Li and D. Thouless, Phys. Rev. B {\bf 40}, 9738 (1989).

\bibitem{BLATT}

V. Geshkenbein, unpublished (1995),
G. Blatter and V. Geshkenbein, in {\it The Physics of Superconductors}, edited by K.H. Bennemann and J.B. Ketterson 
(Springer-Verlag Berlin Heidelberg New York, 2003).

\bibitem{GOROKH}

D.A. Gorokhov and G. Blatter, Phys. Rev. B {\bf 57}, 3586 (1998).

\bibitem{IVLEV1}

B.I. Ivlev, Phys. Rev. A {\bf 62}, 062102 (2000).

\bibitem{IVLEV2}

B.I. Ivlev, Phys. Rev. A {\bf 66}, 012102 (2002).

\bibitem{IVLEV3}

B.I. Ivlev and V. Gudkov, Phys. Rev. C {\bf 69}, 037602 (2004).

\bibitem{IVLEV4}

B.I. Ivlev, Phys. Rev. A {\bf 70}, 032110 (2004).

\bibitem{IVLEV5}

B.I. Ivlev, G. Pepe, R. Latempa, A. Barone, F. Barkov, J. Lisenfeld, and A.V. Ustinov, Phys. Rev. B {\bf 72}, 094507 (2005).
(2005).

\bibitem{ZIMAN}

J.M. Ziman, {\it Principles of the Theory of Solids} (Cambridge University Press, 1964). 

\bibitem{LANDAU1}

L.D. Landau and E. M. Lifshitz, {\it The Classical Theory of Fields} (Butterworth-Heinemann, Oxford, 1998).

\bibitem{IVLEV6}

B.I. Ivlev, arXive:quant-ph/0504206.

\bibitem{WIEL}

W.G. van der Wiel, S. De Franceschi, J.M. Elzerman, T. Fujisava, S. Tarucha, L.P. Kouwenhoven, arXiv:cond-mat/0205350 
(2002). 

\bibitem{LEGGETT}

A.O. Caldeira and A.J. Leggett, Ann. of Phys. {\bf 149}, 374 (1983). 


\end{thebibliography}
\end{document}